\documentclass[a4paper]{article}

\usepackage{spconf,amsmath,graphicx}
\usepackage{amsthm}
\usepackage{algorithm}
\usepackage{algpseudocode}
\usepackage{epstopdf}
\usepackage{amssymb,latexsym}
\usepackage{accents}
\usepackage{caption}
\usepackage{subcaption}
\usepackage{lipsum}
\usepackage[section]{placeins}
\usepackage[noadjust]{cite}
\usepackage{microtype}
\usepackage{lipsum}
\usepackage[framemethod=TikZ]{mdframed}
\newtheorem{rmk}{Remark}

\newcommand{\Q}{{\vphantom{-1}}}

 \newcounter{subeqn} \renewcommand{\thesubeqn}{\theequation\alph{subeqn}}%
 \newcommand{\subeqn}{%
 	\refstepcounter{subeqn}
 	\tag{\thesubeqn}
 }
\begin{document}
\ninept
\title{Secure $\mathit{M}$-PSK Communication via Directional Modulation}
\name{Ashkan Kalantari$^{ \star }$, Mojtaba Soltanalian$^{\dagger }$, Sina Maleki$^{\star }$, Symeon Chatzinotas$^{\star }$, and Bj\"{o}rn Ottersten$^{\star }$
	\thanks{\footnotesize This work was supported by the National Research Fund of Luxembourg under grant number 5798109 and SeMIGod.} 
	\thanks{\footnotesize $^{\star}$  e-mails:\{ashkan.kalantari, sina.maleki, symeon.chatzinotas, bjorn.ottersten\}@uni.lu 
	$^{\dagger }$ e-mail: msol@uic.edu 
		}
	}


\address{$^{\star}$ Interdisciplinary Centre for Security, Reliability and Trust, University of Luxembourg \\
	$^{\dagger}$ Dept. of Electrical and Computer Engineering, University of Illinois at Chicago \\ }
\maketitle
\begin{abstract}
In this work, a directional modulation-based technique is devised to enhance the security of a multi-antenna wireless communication system employing $\mathit{M}$-PSK modulation to convey information. The directional modulation method operates by steering the array beam in such a way that the phase of the received signal at the receiver matches that of the intended $\mathit{M}$-PSK symbol. Due to the difference between the channels of the legitimate receiver and the eavesdropper, the signals received by the eavesdropper generally encompass a phase component different than the actual symbols. As a result, the transceiver which employs directional modulation can impose a high symbol error rate on the eavesdropper without requiring to know the eavesdropper's channel. The optimal directional modulation beamformer is designed to minimize the consumed power subject to satisfying a specific resulting phase and minimal signal amplitude at each antenna of the legitimate receiver. The simulation results show that the directional modulation results in a much higher symbol error rate at the eavesdropper compared to the conventional benchmark scheme, i.e., zero-forcing precoding at the transmitter.    
\end{abstract}
\section{Introduction} \label{sec:intro}
Due to the broadcast nature of wireless communications, sensitive information can be exposed to unintended receivers. A recent effort in order to protect the information in the physical layer has been carried out by relying on the information-theoretic concept introduced in~\cite{Wyner:1975}. This type of coding helps achieving a specific rate, known as the secrecy rate, with which the transmission is completely secure. Secrecy rate was later extended to broadcast~\cite{Csiszar:1978}, Gaussian~\cite{Leung-Yan-Cheong:1978}, and fading channels~\cite{Gopala:2008,Zang:2007,Oggier:2008}. One drawback is that to calculate the secrecy rate, channel state information (CSI) of the eavesdropper is required, which is difficult to get in practice, specially for a passive eavesdropper. 

We further note that many communication systems use finite-alphabet signals; in particular, $\mathit{M}$-PSK modulation has various applications in wireless networks~\cite{MAC:standard:2004}, ZigBee protocol~\cite{zig:stan:2007} and multi-user communications~\cite{cons:2015,Masouros:2015}. Since finite-alphabet signals usually have a non-Gaussian distribution~\cite{MADISETTI1997}, they are not optimal in terms of the developed secrecy rates in~\cite{Wyner:1975,Csiszar:1978,Leung-Yan-Cheong:1978,Gopala:2008,Zang:2007,Oggier:2008}. There has been research interest on the security improvement when finite-alphabet signals are used. The authors in~\cite{Bakr:2010} devote some of the available power to add a randomly scaled version of the finite-alphabet signal to the signal itself without optimal beamforming. If the added random part rotates the $\mathit{M}$-PSK constellation enough, the eavesdropper detects the wrong symbol. In~\cite{Xiaohua:2007}, suboptimal random beamforming is used to enhance the security without requiring the eavesdropper CSI when finite-alphabet signal is used. An external helper generating interference in the form of fine-alphabet signal is considered in~\cite{Chorti:2012:main}. Information-theoretic secrecy rate expressions are derived by approximating the helping interference distribution as sum of the Gaussian distributions and assuming the availability of the eavesdropper's CSI. The authors in~\cite{Bashar:2012} study the information-theoretic secrecy rate for finite-alphabet signals in a communication system with multi-antenna nodes by assuming the eavesdropper CSI availability at the transmitter. In another paradigm in~\cite{Valliappan:2013}, random and optimized antenna subset selection from a large uniform linear antenna array system without optimal beamforming is employed in order to improve the security in a milliliter-wave system with line-of-sight channel. 

In this work, rather than relying on the information-theoretic security concept of~\cite{Wyner:1975}, a signal processing approach based on the array-based directional modulation~\cite{Daly:2009,Daly:2010} concept is utilized in order to enhance the security, without requiring the eavesdropper's CSI, when $\mathit{M}$-PSK modulation is used for communication. In this technique, instead of producing the symbols at the transmitter, the phase and amplitude of each element of the array is adjusted so that the resulting phase of the received signals on each antenna of the receiver is equal to the phase of a specific $\mathit{M}$-PSK symbol. We assume the eavesdropper channel is independent from the one of the legitimate receiver. Therefore, the received signals by the eavesdropper have a different resulting phase compared to the legitimate receiver. As shall be shown later, this increases the symbol error rate (SER) at the eavesdropper considerably.

The summarized contributions are as follows. The optimal beamformer for a multi-antenna transmitter is designed when directional modulation is used for $\mathit{M}$-PSK transmission. There are several works such as~\cite{Tekin:2008,Zhu:2009,Tie:2009,Dong:2010,Rongqing:2010,Jing:2010,Yuksel:2011,Hang:2013,Kalantari:2015:TWC,Kalantari:2015} which perform secrecy rate analysis by assuming the availability of the eavesdropper's CSI, which is difficult to acquire in practice. However, in the directional modulation, the security is enhanced without requiring the eavesdropper CSI while assuming that the eavesdropper is aware of the global CSI as well as the transmitter and receiver configurations, including the number of antennas and the modulation order. Although the information-theoretic secrecy rate provides perfect secrecy, i.e., zero bit leakage, it reduces the message transmission rate. Here, we rely on a signal processing-based approach, so the transmission rate does not need to be scarified in order to enhance the security. Finally, the transmission channel CSI is not required at the legitimate receiver when using the directional modulation, in addition, there is no need for zero-forcing (ZF) or minimum-mean-square-error (MMSE) multi-input and multi-output (MIMO) receiver implementation at the legitimate receiver side. 

The remainder of the paper is organized as follows. In Section~\ref{sec:sys:mod}, the network configuration as well as the signal models are introduced. The security and the beamformer design for the directional modulation is mentioned in Section~\ref{sec:phy:sec}. In Section~\ref{sec:sim}, the security of the directional modulation is evaluated using simulations. Finally, the conclusions are drawn in Section~\ref{sec:con}. 

\emph{Notation}: Upper-case and lower-case bold-faced letters are used to denote matrices and column vectors, respectively. Superscripts $(\cdot)^T$, $(\cdot)^*$, $(\cdot)^H$ represent transpose, conjugate, and Hermitian operators, respectively. ${{\bf{I}}_{N \times N}}$ denotes an $N$ by $N$ identity matrix, $diag(\bf{a})$ denotes a diagonal matrix where the elements of $\bf{a}$ are its diagonal entries, ${\bf{a}} \circ {\bf{b}}$ is the element-wise Hadamard product, $\|\cdot\|$ is the Frobenius norm, and $|\cdot|$ represents the absolute value of a scalar. ${\mathop{\rm Re}\nolimits} \left(  \cdot  \right)$, ${\mathop{\rm Im}\nolimits} \left(  \cdot  \right)$, and $\arg \left(  \cdot  \right)$ represent the real part, imaginary part, and angle of a complex number, respectively.
\section{Signal and System Model} \label{sec:sys:mod}
\begin{figure}[t!]
	\centering
	\includegraphics[width=8cm]{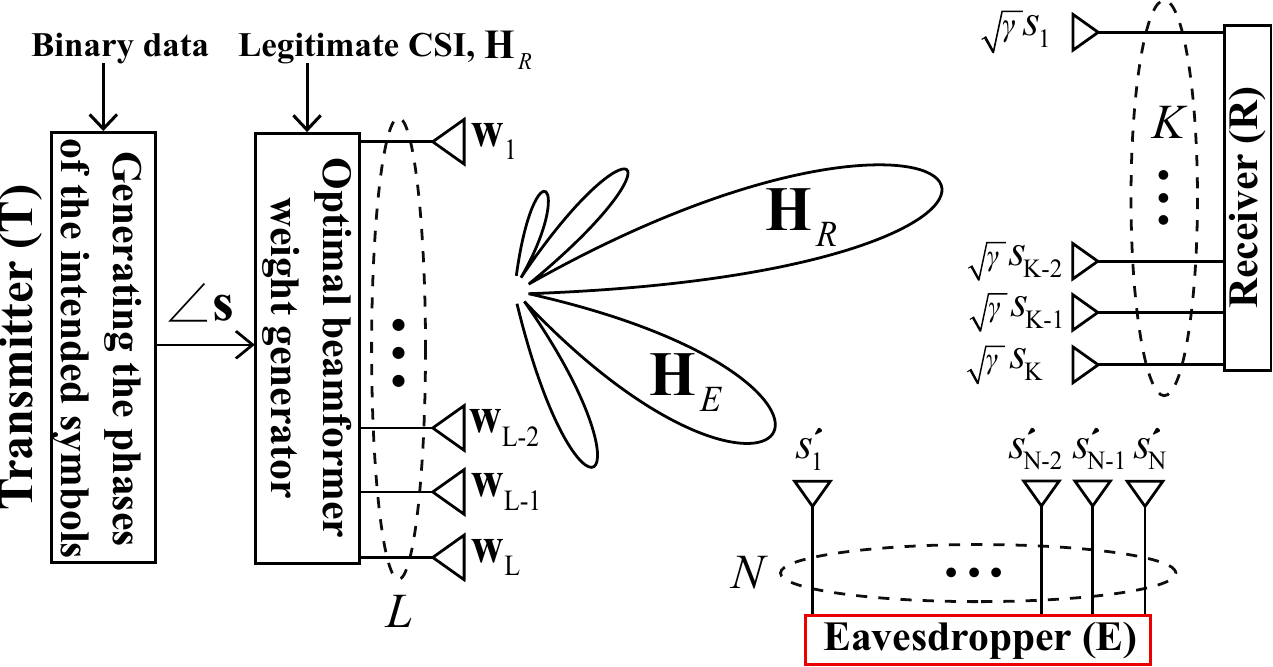}
	\caption{Array-based directional modulation to enhance the security in a MIMO wiretap channel.}
	\label{fig:sys:mod}
\end{figure}
We consider a communication network with a multi-antenna transmitter denoted by $T$, a multi-antenna receiver denoted by $R$, and a multi-antenna eavesdropper denoted by $E$ as shown in Fig.~\ref{fig:sys:mod}. Using the transmission channel CSI and the symbols' phases, the phase and the amplitude of each transmit antenna is designed so that the resulting phase of the received signals by each antenna of the legitimate receiver is equal to the phase of a specific $\mathit{M}$-PSK symbol. Here, all the communication channels are considered to be quasi-static block fading. After applying the optimal coefficients to array elements, the received signals at $R$ and $E$ are
\begin{align}
&{{\bf{y}}_{{R}^\Q}} = {\bf{H}}_R{\bf{w}} + {{\bf{n}}_R^\Q}, 
\\
&{{\bf{y}}_{E}^\Q} = {\bf{H}}_E{\bf{w}} + {{\bf{n}}_E^\Q},
\label{eqn:eve:rec:dm}
\end{align}
where the random variables ${{\bf{n}}_R^\Q}$ and ${{\bf{n}}_E^\Q}$ denote the additive white Gaussian noise at $R$ and $E$, respectively. The Gaussian random variables ${{\bf{n}}_R^\Q}$ and ${{\bf{n}}_E^\Q}$ are independent and identically distributed (i.i.d.) with ${{\bf{n}}_R^\Q}\sim\mathcal{CN}({\bf{0}}, \sigma _{{n_R}}^2  {{\bf{I}}_{K \times K}} )$, and ${{\bf{n}}_E^\Q}\sim\mathcal{CN}({\bf{0}},  \sigma _{{n_E}}^2  {{\bf{I}}_{N \times N}} )$, respectively, where $\mathcal{CN}$ denotes the complex and circularly symmetric i.i.d. random variable. The signal ${{\bf{y}}_R^\Q}$ is a $K \times 1$ vector denoting the received signals by $R$, ${{\bf{y}}_E^\Q}$ is an $N \times 1$ vector denoting the received signals by $E$, ${{\bf{H}}_R}$ is a $K \times L$ matrix denoting the channel from $T$ to $R$ defined as ${{\bf{H}}_R} = {[{{\bf{h}}_1},...,{{\bf{h}}_i},...,{{\bf{h}}_R}]^T}$, ${{\bf{H}}_E}$ is an $N \times L$ matrix denoting the channel from $T$ to $E$, and $\bf{w}$ is the beamforming vector. In the directional modulation scheme, the elements of the vector ${\bf{H}}_R{\bf{w}}$ are the $\mathit{M}$-PSK symbols. Hence, to detect the received symbols, $R$ can directly apply a conventional detector, e.g., maximum-likelihood (ML) detector, on each individual element of the vector ${\bf{y}}_{R}^\Q$ without requiring to implement a ZF or MMSE receiver~\cite{Biglieri2007}.
\section{Security via directional modulation}   \label{sec:phy:sec}
In this section, first, the security benefits of the directional modulation is discussed, and then the optimal beamformer design problem is formulated and solved.
\subsection{Security Advantages}   \label{subsec:sec:adv}
As indicated earlier, the design of the beamformer in the directional modulation scheme is based on the phase of the symbols which are to be conveyed to $R$; in particular, to recover the symbols, it is necessary for $E$ to somehow estimate ${{\bf{H}}_R}{\bf{w}}$. To this end, considering that $E$ is aware of ${{\bf{H}}_R}$, $E$ should begin by estimating $\bf{w}$ given by
\begin{align}
\widehat {\bf{w}}  = {\left( {{\bf{H}}_E^H{{\bf{H}}_E}} \right)^{ - 1}}{\bf{H}}_E^H{{\bf{y}}_E} = {\bf{w}} + {\left( {{\bf{H}}_E^H{{\bf{H}}_E}} \right)^{ - 1}}{\bf{H}}_E^H{{\bf{n}}_E^\Q},
\label{eqn:eve:zf}
\end{align}
where $\widehat {\bf{w}}$ is the estimation of $\bf{w}$ at $E$. Note that $E$ can compute~\eqref{eqn:eve:zf} only in the case $N \ge L$ since ${\left( {{\bf{H}}_E^H{{\bf{H}}_E}} \right)^{-1}}{\bf{H}}_E^H{{\bf{H}}_E} \ne {\bf{I}}$ for $N<L$. Next, $E$ needs to multiply $\widehat {\bf{w}}$ by ${\bf{H}}_R$ to calculate ${{\bf{H}}_R}{\bf{w}}$, viz.
\begin{align}
{\bf{H}}_R \widehat {\bf{w}} = {\bf{H}}_R{\bf{w}} + {\bf{H}}_R{\left( {{\bf{H}}_E^H{{\bf{H}}_E}} \right)^{ - 1}}{\bf{H}}_E^H{{\bf{n}}_E^\Q}.
\label{eqn:eve:mim}
\end{align}
where ${\bf{H}}_R \widehat {\bf{w}}$ denotes the estimation of ${{\bf{H}}_R}{\bf{w}}$ at $E$. However, the first step of estimating ${{\bf{H}}_R}{\bf{w}}$ in~\eqref{eqn:eve:zf} results in the noise enhancement at $E$~\cite{Arokiamary2009}. Therefore, the SER at $E$ will be higher than that of $R$, especially in a low signal-to-noise ratio regime due to the fact that $R$ can directly detect the $\mathit{M}$-PSK symbols without channel equalization. Moreover, in the case of $N<L$, $E$ cannot estimate ${{\bf{H}}_R}{\bf{w}}$. Therefore, $E$ can not correctly decode the transmitted signal, and any attempt to do so leads to a high SER. 
\begin{rmk}
It is interesting to observe that the condition $N<L$ is easily met in a massive MIMO scenario, hence, the directional modulation technique appears to be a good candidate for security enhancement in massive MIMO systems. \qquad \qquad \qquad \qquad \,\,\,\,\,\,\,\,\,\,\, $ \blacksquare $ 
\end{rmk}
In the next section, the optimal beamformer design problem for the directional modulation is formulated and solved from a power efficiency viewpoint.
\subsection{Optimal Beamformer Design}\label{subsec:opt:bea}
In this section, the optimal beamformer design problem for the directional modulation is defined and transformed into a linearly constrained quadratic program which can be solved efficiently. Herein, the beamformer for the directional modulation will be designed to minimize the consumed power at the transmitter such that 1) the resulting phase of the signals received by each antenna of $R$ is equal to the phase of a specific $\mathit{M}$-PSK symbol, and that 2) the required signal level for the in-phase and quadrature-phase components of the resulting $\mathit{M}$-PSK symbol on each antenna of $R$ is preserved above a specific level. Note that in such a setup, by minimizing the power we actually increase the SER at $E$ even when $N \ge L$ while keeping the quality of our own signal reception at the desired level.

Using the directional modulation signal model described in Section~\ref{sec:sys:mod}, the related beamformer design problem can be cast as
\begin{subequations}
\begin{align}
&\mathop {\min }\limits_{\bf{w}} {\mkern 1mu} {\mkern 1mu} {\mkern 1mu} {\mkern 1mu} {\left\| {\bf{w}} \right\|^2}
\nonumber\\
& \,\, \text{s.t.}   \,\,\,\,   \arg \left( {{{\bf{h}}_i^T}{\bf{w}}} \right) = \arg \left( {{s_i}} \right),  \qquad \,\,\,\,\,\,\,\,\,\,\,\,\,\,\,\,\,  \forall \,\, i  
\subeqn \label{subeq:c1}
\\
&                  \qquad \,\,  {\rm{Re}}\left( {{s_i}} \right){\rm{Re}}\left( {{{\bf{h}}_i^T}{\bf{w}}} \right) \ge \sqrt \gamma {\mkern 1mu} {{\mathop{\rm Re}\nolimits} ^2}\left( {{s_i}} \right),  \,\, \forall \,\, i 
\subeqn \label{subeq:c2}
\\
&                  \qquad \,\,  {\rm{Im}}\left( {{s_i}} \right){\rm{Im}}\left( {{{\bf{h}}_i^T}{\bf{w}}} \right) \ge \sqrt \gamma {\mkern 1mu} {{\mathop{\rm Im}\nolimits} ^2}\left( {{s_i}} \right),  \,\, \forall \,\, i
\subeqn \label{subeq:c3}
\end{align}
\label{eqn:opt:dm}%
\end{subequations}
where $s_i$ is the $i$-th $\mathit{M}$-PSK symbol possessing instantaneous unit energy, i.e., ${\left| s_i \right|^2} = 1$, and $\sqrt \gamma$ is a scalar to adjust the required level for the in-phase and quadrature-phase components of the received signal at the corresponding antenna of $R$. Note that since the in-phase or quadrant-phase part of the symbol may be negative, both sides of the constraints~\eqref{subeq:c2} and~\eqref{subeq:c3} are multiplied by ${\rm Re}(s_i)$ and ${\rm Im}(s_i)$, respectively. Since~\eqref{subeq:c1} holds at the optimal point, ${\rm Re}(s_i)$ and ${\rm Im}(s_i)$ have the same sign as ${\mathop{\rm Re}\nolimits} \left( {{\bf{h}}_i^T{\bf{w}}} \right)$ and ${\mathop{\rm Im}\nolimits} \left( {{\bf{h}}_i^T{\bf{w}}} \right)$ at the optimal point. As a result, the multiplication at both sides of~\eqref{subeq:c2} and~\eqref{subeq:c3} does not change the side of the inequality. 

To simplify~\eqref{eqn:opt:dm}, let's write the constraint~\eqref{subeq:c1} as 
\begin{align}
{\mathop{\rm Re}\nolimits} \left( {{{\bf{h}}_i^T}{\bf{w}}} \right){\alpha _i} 
- {\mathop{\rm Im}\nolimits} \left( {{{\bf{h}}_i^T}{\bf{w}}} \right) = 0, \,\,\, i=1,...,K,
\label{eqn:pha}
\end{align}
where ${\alpha _i} = \tan \left( {{s_i}} \right)$. Using the equations derived in~\eqref{eqn:pha} and by putting together the constraints ~\eqref{subeq:c2} and~\eqref{subeq:c3}, it is possible to reformulate~\eqref{eqn:opt:dm} into a compact form as
\begin{subequations}
\begin{align}
& \mathop {\min }\limits_{\bf{w}}  \,\,\,\,   {\left\| {\bf{w}} \right\|^2}
\nonumber\\
& \,\, \text{s.t.}   \,\,\,\,\,\,\,\,  {\bf{A}}{\rm{Re}}\left( {{{\bf{H}}_R}{\bf{w}}} \right) - {\rm{Im}}\left( {{{\bf{H}}_R}{\bf{w}}} \right) = {\bf{0}},
\label{subeq:c21}
\\
&                  \qquad \,\,\,\,\,   {\mathop{\rm Re}\nolimits} \left( {\bf{S}} \right){\mathop{\rm Re}\nolimits} \left( {{{\bf{H}}_R}{\bf{w}}} \right) \ge {\sqrt \gamma} \,  {{\bf{s}}_r},
\label{subeq:c22}
\\
&                  \qquad \,\,\,\,\,   {\mathop{\rm Im}\nolimits} \left( {\bf{S}} \right) {\rm Im}\left( {{{\bf{H}}_R}{\bf{w}}} \right)\ge {\sqrt \gamma} \,  {{\bf{s}}_i},
\label{subeq:c23}
\end{align}
\label{eqn:opt:dm:sim}%
\end{subequations}
where ${\bf{S}} = diag\left( {\bf{s}} \right)$, ${\bf{s}} = {\left[ {{s_1},...,s_i,...,{s_K}} \right]^T}$ is the vector bearing the $\mathit{M}$-PSK symbols, ${{\bf{s}}_r} = {\mathop{\rm Re}\nolimits} \left( {\bf{s}} \right) \circ {\mathop{\rm Re}\nolimits} \left( {\bf{s}} \right)$, ${{\bf{s}}_i} = {\mathop{\rm Im}\nolimits} \left( {\bf{s}} \right) \circ {\mathop{\rm Im}\nolimits} \left( {\bf{s}} \right)$, and ${\bf{A}} = diag\left( {{\alpha _1},...,{\alpha _K}} \right)$. 
 
Note that applying $tan(\cdot)$ on the phases of the intended symbols causes ambiguity since symbols with different phases can have the same $tan$ value, e.g., $\tan \left( {\frac{\pi }{4}} \right) = \tan \left( {\frac{{3\pi }}{4}} \right)$. Hence, the constraints ${\mathop{\rm Re}\nolimits} \left( {\bf{S}} \right){\mathop{\rm Re}\nolimits} \left( {{{\bf{H}}_R}{\bf{w}}} \right) \ge 0$ and ${\mathop{\rm Im}\nolimits} \left( {\bf{S}} \right) {\rm Im}\left( {{{\bf{H}}_R}{\bf{w}}} \right) \ge 0$ need to be added to the design problem~\eqref{eqn:opt:dm:sim} to resolve the phase ambiguity. Interestingly, these constraints are already present in~\eqref{subeq:c22} and~\eqref{subeq:c23}.
 
To transform~\eqref{eqn:opt:dm:sim} into a familiar form, we represent ${{\bf{H}}_R} = {\rm{Re}}\left( {{{\bf{H}}_R}} \right) + i{\rm{Im}}\left( {{{\bf{H}}_R}} \right)$ and ${\bf{w}} = {\mathop{\rm Re}\nolimits} \left( {\bf{w}} \right) + i{\mathop{\rm Im}\nolimits} \left( {\bf{w}} \right)$ in order to write ${\bf{H}}_R{\bf{w}}$ as 
 \begin{align}
 {{\bf{H}}_R}{\bf{w}} =& {\rm{Re}}\left( {{{\bf{H}}_R}} \right){\rm{Re}}\left( {\bf{w}} \right) - {\rm{Im}}\left( {{{\bf{H}}_R}} \right){\rm{Im}}\left( {\bf{w}} \right)
 \nonumber\\
 &+ i\left[ {{\rm{Re}}\left( {{{\bf{H}}_R}} \right){\rm{Im}}\left( {\bf{w}} \right) + {\rm{Im}}\left( {{{\bf{H}}_R}} \right){\rm{Re}}\left( {\bf{w}} \right)} \right],
 \label{eqn:Hw}
 \end{align}
 which helps us to write the real and imaginary parts of ${\bf{H}}_R{\bf{w}}$ as
 \begin{align}
 {\rm{Re}}\left( {{\bf{H}}_R{\bf{w}}} \right) = {\bf{H}}_{R_1}\widetilde {\bf{w}}, \,\, {\rm{Im}}\left( {{\bf{H}}_R{\bf{w}}} \right) = {\bf{H}}_{R_2}\widetilde {\bf{w}},
 \label{eqn:rel:ima:hw}
 \end{align}
 where $\widetilde {\bf{w}} = {\left[ {{\rm{Re}}\left( {\bf{w}}^T \right),{\rm{Im}}\left( {\bf{w}}^T \right)} \right]^T}$, ${{\bf{H}}_{{R_1}}} = \left[ {{\mathop{\rm Re}\nolimits} \left( {{{\bf{H}}_R}} \right), - \rm Im\left( {{{\bf{H}}_R}} \right)} \right]$, ${{\bf{H}}_{{R_2}}} = \left[ {\rm Im\left( {{{\bf{H}}_R}} \right),{\mathop{\rm Re}\nolimits} \left( {{{\bf{H}}_R}} \right)} \right]$. Also, it is straightforward to see that ${\left\| {\widetilde {\bf{w}}} \right\|^2} = {\left\| {\bf{w}} \right\|^2}$. 
 
 Using the derivations in~\eqref{eqn:rel:ima:hw},~\eqref{eqn:opt:dm:sim} can be reformulated as
\begin{align}
& \mathop {\min }\limits_{\widetilde {\bf{w}}}  \,\,\,\,  \left\| {\widetilde {\bf{w}}} \right\|^2
\nonumber\\
& \, \text{s.t.}   \,\,\,\,\,\,\,  \left( {{\bf{A}}{{\bf{H}}_{{R_1}}} - {{\bf{H}}_{{R_2}}}} \right)\widetilde {\bf{w}} =0,
\nonumber\\
&                  \qquad \,\,\,\,   {\mathop{\rm Re}\nolimits} \left( {\bf{S}} \right){{\bf{H}}_{{R_1}}}\widetilde {\bf{w}} \ge {\sqrt \gamma} \,  {{\bf{s}}_r},
\nonumber\\
&                  \qquad \,\,\,\,  {\mathop{\rm Im}\nolimits} \left( {\bf{S}} \right){{\bf{H}}_{{R_2}}}\widetilde {\bf{w}} \ge {\sqrt \gamma} \,  {{\bf{s}}_i}.
\label{eqn:opt:dm:sim:2}
\end{align}
For~\eqref{eqn:opt:dm:sim:2} to be feasible, $\widetilde {\bf{w}}$ should lie in the null space of the matrix ${{\bf{A}}{{\bf{H}}_{{R_1}}} - {{\bf{H}}_{{R_2}}}}$. If the singular value decomposition of ${{\bf{A}}{{\bf{H}}_{{R_1}}} - {{\bf{H}}_{{R_2}}}}$ is shown by ${\bf{U}}\Sigma {{\bf{V}}^H}$, the orthonormal basis for the null space of ${{\bf{A}}{{\bf{H}}_{{R_1}}} - {{\bf{H}}_{{R_2}}}}$ are the last $2L - K$ columns of $\bf{V}$ which span $\widetilde {\bf{w}}$~\cite{Strang2009}. This means that the vector $\widetilde {\bf{w}}$ can be written as $\widetilde {\bf{w}} = {\bf{E \boldsymbol \lambda }}$ where ${\bf{E}} = \left[ {{{\bf{v}}_{K + 1}},...,{{\bf{v}}_{2L}}} \right]$ and $\boldsymbol \lambda  = \left[ {{\lambda _1},...,{\lambda _{2L-K}}} \right]$. Therefore,~\eqref{eqn:opt:dm:sim:2} boils down into
\begin{align}
& \mathop {\min }\limits_{ {\boldsymbol{\lambda}}}  \,\,\,\,  \left\| { {{\boldsymbol{\lambda }}}} \right\|^2
\nonumber\\
& \, \text{s.t.}   \,\,\,\,\,\,\,\,  {\mathop{\rm Re}\nolimits} \left( {\bf{S}} \right){{\bf{H}}_{{R_1}}} {\bf{E}}  {\boldsymbol{\lambda}} \ge {\sqrt \gamma} \,  {{\bf{s}}_r},
\nonumber\\
&                  \qquad \,\,\,\,   {\mathop{\rm Im}\nolimits} \left( {\bf{S}} \right){{\bf{H}}_{{R_2}}} {\bf{E}} {\boldsymbol{\lambda}} \ge {\sqrt \gamma}  \,  {{\bf{s}}_i},
\label{eqn:opt:dm:fin}
\end{align}
which is a convex linearly constrained quadratic programming and can be solved efficiently using standard convex optimization techniques. 
\begin{rmk}
Since $\widetilde {\bf{w}}$ in~\eqref{eqn:opt:dm:sim:2} is spanned by the last $2L - K$ vectors of the matrix $\bf{V}$, a necessary condition for the existence of the optimal beamformer for the directional modulation is $L > \frac{K}{2}$ which means that the number of transmit antennas needs to be more than half of the number of antennas at the legitimate receiver. Provided that the latter condition is met, a sufficient condition can be proposed from a geometrical point of view; namely that the feasible set of~\eqref{eqn:opt:dm:fin} is not empty if and only if the intersection of the linear spaces in the constraint set constitutes a non-empty set. \qquad \qquad \qquad \qquad \,\,\,\,\,\,\,\, $ \blacksquare $ 
\end{rmk}
\section{Simulation Results}   \label{sec:sim}
In this section, the performance of the secure directional modulation and a benchmark scheme are demonstrated and compared using different simulation scenarios. In all simulations, channels are considered to be quasi static block Rayleigh fading which are generated using i.i.d. complex Gaussian random variables with distribution$\sim\mathcal{CN}(0 , \sigma^2 )$ and remain fixed during the interval in which the $\mathit{M}$-PSK symbol is being conveyed to $R$. In addition, the noise is also generated using i.i.d. complex Gaussian random variables, and the modulation order used in all of the scenarios is $8$-PSK. The acronym ``DM'' is used instead of the term ``directional modulation'' in the legend of the figures. Before proceeding, we first mention the benchmark scheme.
\begin{figure}[]
	\centering
	\includegraphics[width=7.5cm]{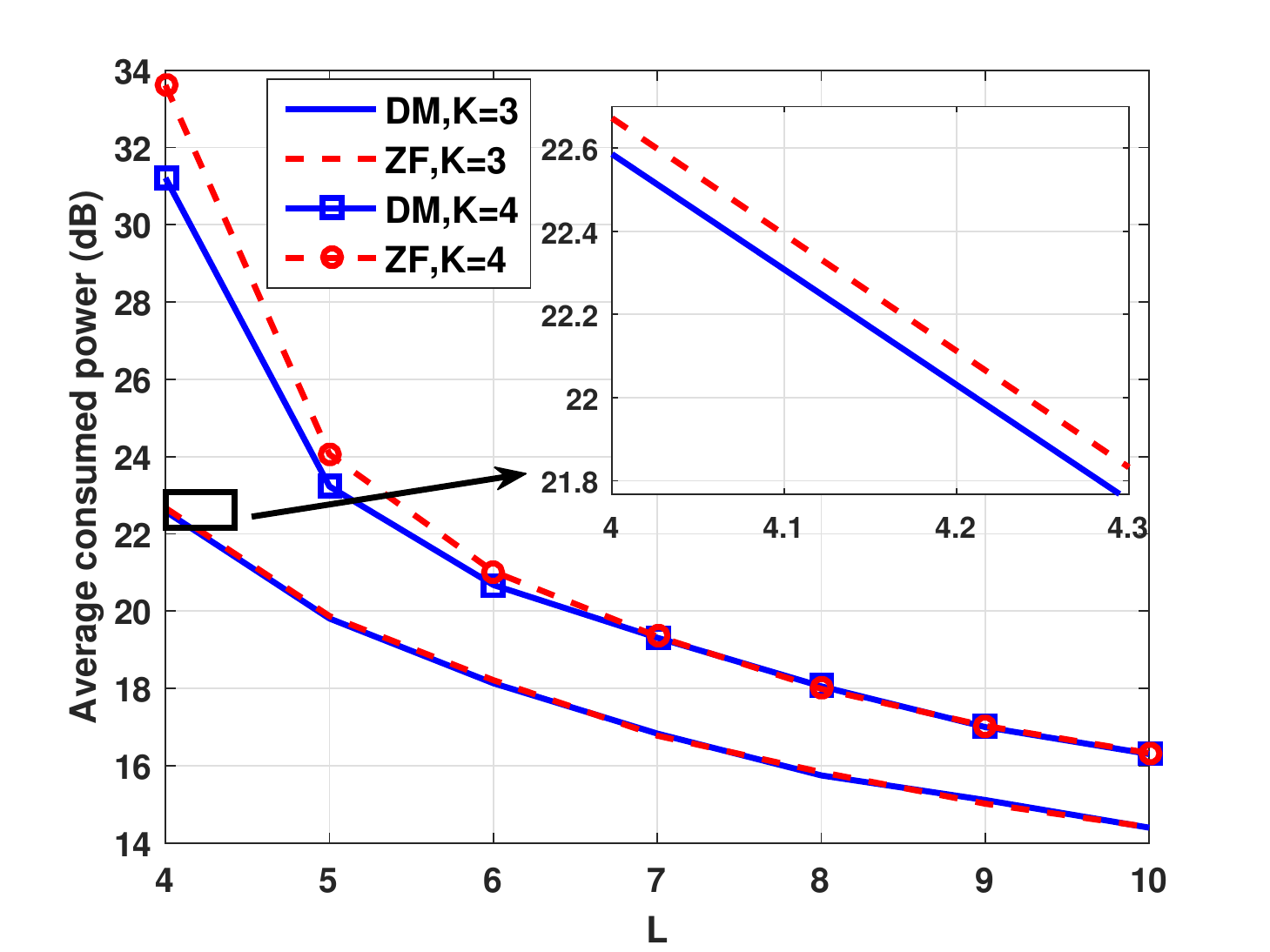}
	\caption{Average consumed power with respect to the number of transmitting antennas for the directional modulation and the benchmark schemes when $N=6$, $\sqrt \gamma=8$, and $\beta=8$.}
	\label{fig:pow:ant}
\end{figure}

The ZF at the transmitter in order to neutralize the interference between received symbol streams at $R$~\cite{Lai-U:2004} is used as the comparison benchmark. In contrast to the directional modulation scheme, the symbols in the benchmark scheme are generated and sent from the transmitter. The received signals at $R$ and $E$ are
\begin{align}
&{{\bf{y}}_R^\Q} = {{\bf{H}}_R}{\bf{Ps}} + {{\bf{n}}_R^\Q}, 
\\
&{{\bf{y}}_E^\Q} = {{\bf{H}}_E}{\bf{Ps}} + {{\bf{n}}_E^\Q},
\label{eqn:rec:sig:zf}
\end{align}
where $\bf{P}={\bf{H}}_R^H{\left( {{{\bf{H}}_R}{\bf{H}}_R^H} \right)^{ - 1}} \beta$, and $\beta$ is the amplification factor for the symbols which acts similar as $\sqrt \gamma$ in the directional modulation scheme. After the signal is received by $E$, it estimates the transmitted symbols as 
\begin{align}
\widehat {\bf{s}} &= {\left[ {{{\left( {{{\bf{H}}_E}{\bf{P}}} \right)}^H}{{\bf{H}}_E}{\bf{P}}} \right]^{ - 1}}{\left( {{{\bf{H}}_E}{\bf{P}}} \right)^H}{{\bf{y}}_E}
\nonumber\\
&= {\bf{s}} + {\left[ {{{\left( {{{\bf{H}}_E}{\bf{P}}} \right)}^H}{{\bf{H}}_E}{\bf{P}}} \right]^{ - 1}}{\left( {{{\bf{H}}_E}{\bf{P}}} \right)^H}{{\bf{n}}_E}.
\label{eqn:eve:rec:sig:zf}
\end{align}
The dimension of the matrix ${{\bf{H}}_E}{\bf{P}}$ is $N \times K$ which results in ${\left[ {{{\left( {{{\bf{H}}_E}{\bf{P}}} \right)}^H}{{\bf{H}}_E}{\bf{P}}} \right]^{ - 1}}{\left( {{{\bf{H}}_E}{\bf{P}}} \right)^H}{{\bf{H}}_E}{\bf{P}}=\bf{I}$ for $N \ge K$. This means that $E$ can recover the symbols even with the condition $N<L$. Generally, satisfying the condition $N<L$ is easier than $N<K$ since the base station has usually more antennas than the users. Considering that the condition for $E$ to estimate ${{\bf{H}}_R}{\bf{w}}$ and detect the $\mathit{M}$-PSK symbol in the directional modulation scheme is $ N \ge L$, the directional modulation is much more probable to enhance the security compared to the benchmark scheme.

In the first scenario, the effect of the number of transmission antennas, $L$, on the consumed power is investigated. The average consumed power with respect to $L$ is shown in Fig.~\ref{fig:pow:ant}. As seen, for a specific range of $L$, the directional modulation consumes less power than the benchmark scheme. Furthermore, the difference in power consumption increases when the number of antennas at $R$ increases.

\begin{figure}[]
	\centering
	\includegraphics[width=7.5cm]{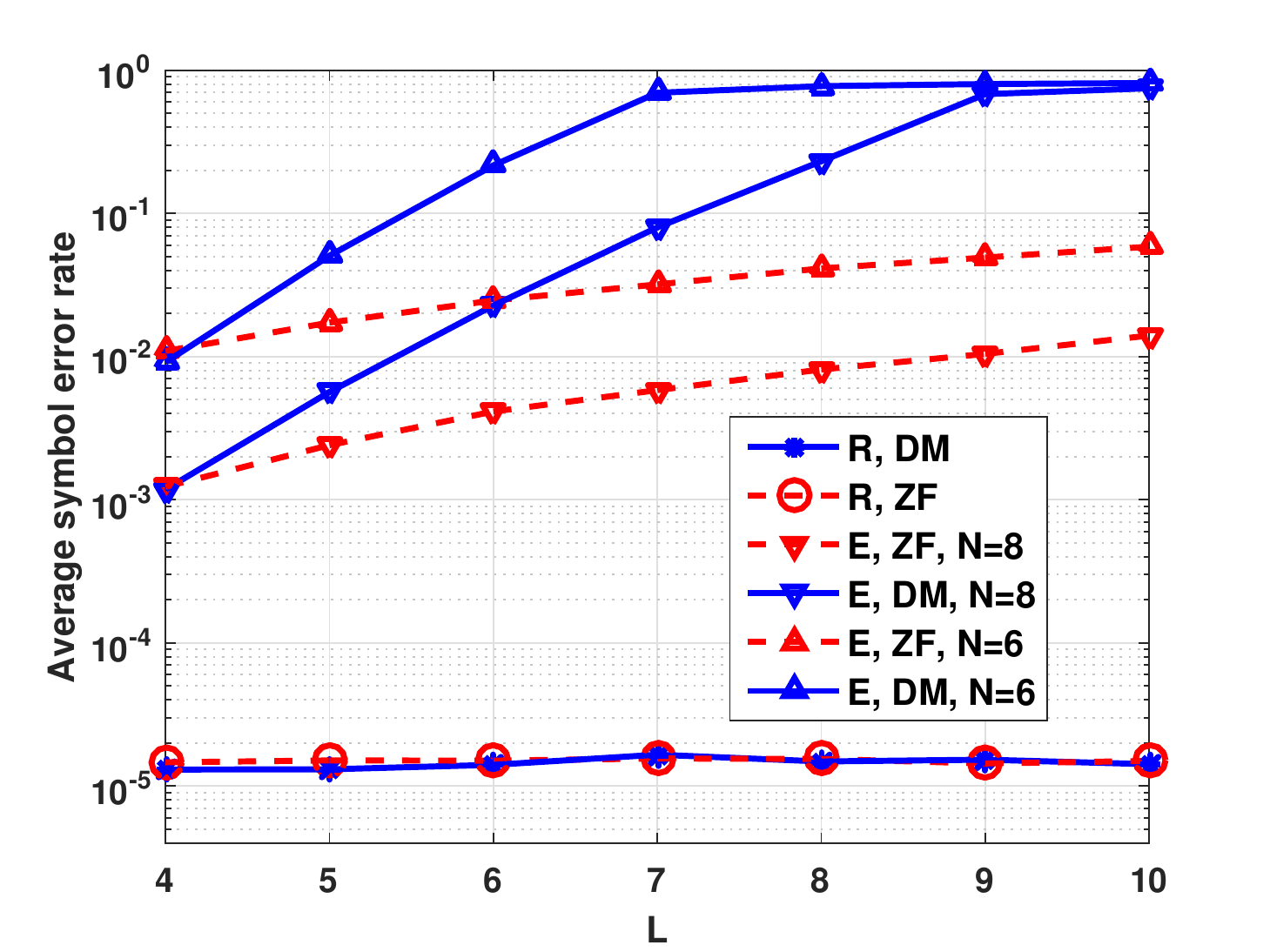}
	\caption{Average SER versus the number of transmitting antennas for the directional modulation and the benchmark schemes when $K=4$, $\sqrt \gamma=8$, and $\beta=8$.}
	\label{fig:ser:ant}
\end{figure}

The SER at $R$ and $E$ when using different number of transmitting antennas is studied in the second scenario. The average SER with respect to $L$ is presented in Fig.~\ref{fig:ser:ant}. As it is seen, the directional modulation causes considerably more SER at $E$ compared to the benchmark scheme. Furthermore, as the antennas of $E$ increase, the difference between the SER caused at $E$ by the directional modulation and benchmark scheme increases for specific values of $L$. For example, when $L=9$ and $N=8$, the difference between the SER caused by the directional modulation and ZF schemes is more than the case when $L=9$ and $N=6$. 
\begin{figure}[]
	\centering
	\includegraphics[width=7.5cm]{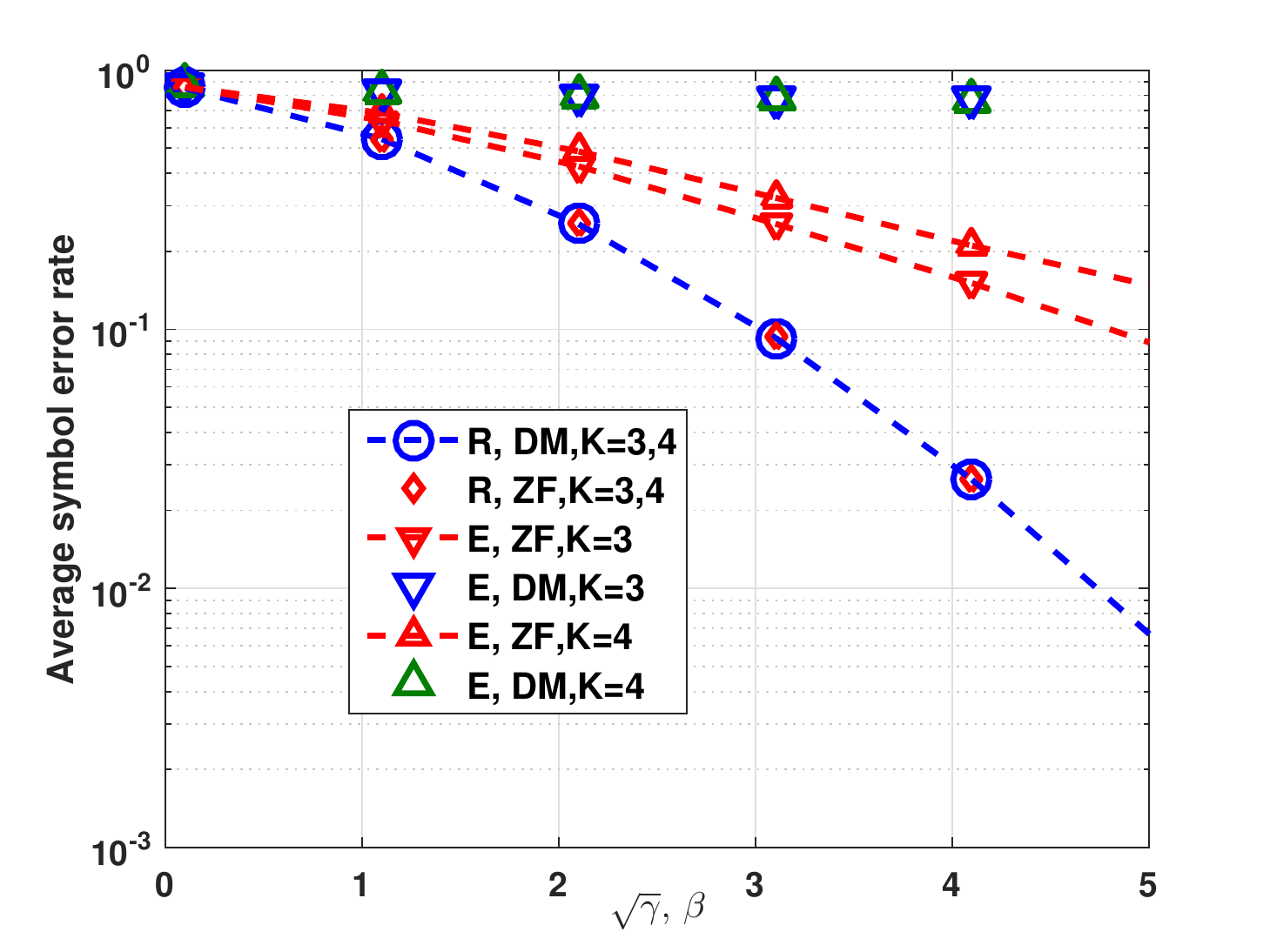}
	\caption{Average SER versus the in-phase and the quadrature-phase required signal levels for directional modulation and benchmark schemes when $L=8$ and $N=6$.}
	\label{fig:ser:snr}
\end{figure}

In the last scenario, the relation between the required level, $\sqrt \gamma$, for the in-phase and quadrature-phase components of the induced symbols at $R$ and the SER at $R$ and $E$ is studied. The average SER at $R$ and $E$ with respect to $\sqrt \gamma$ is shown in Fig.~\ref{fig:ser:snr}. As it is observed, when using the benchmark scheme, the SER at $E$ decreases as $\sqrt \gamma$ increases. This reduction is more when the antennas of $R$ decreases. On the other hand, when using the directional modulation scheme, the SER at $E$ does not decrease as the required signal level at the legitimate increases. As mentioned, this is due to the fact that for $N \ge K$, $E$ can remove the effect of precoder in the benchmark scheme and decrease its own SER. However, as explained in Section~\ref{subsec:sec:adv}, $E$ cannot estimate ${{\bf{H}}_R}{\bf{w}}$ when $N<L$ in the directional modulation. Therefore, assuming the independence of ${{\bf{H}}_R}$ and ${{\bf{H}}_E}$, $E$ has to detect the symbols according to the phases of the vector ${{\bf{H}}_E}{\bf{w}}$ which are different from the phases of the vector ${{\bf{H}}_R}{\bf{w}}$. 
\section{Conclusions}\label{sec:con}
The optimal beamformer for the secure directional modulation was designed without requiring the eavesdropper's CSI. It was seen that the eavesdropper cannot regenerate the beamformer and has to estimate the signal received by the legitimate receiver using the global CSI knowledge. However, this estimation enhances the noise and is only possible when the transmitter has less antennas than the eavesdropper. The directional modulation was compared with the ZF at the transmitter as the benchmark. In the ZF scheme, the eavesdropper had to have more antennas than the legitimate receiver to recover the symbol. The results showed that directional modulation leads into less power consumption and more SER at the eavesdropper compared to the conventional benchmark scheme. These observations confirm the reliability of the studied directional modulation approach from a security point of view.  

\end{document}